\documentclass[%
reprint,
superscriptaddress,
amsmath,amssymb,aps,
]{revtex4-2}

\usepackage{color}
\usepackage{graphicx}
\usepackage{dcolumn}
\usepackage{bm}
\usepackage{caption} 
\usepackage{subcaption} 
\usepackage{amsthm}
\theoremstyle{plain}

\newtheorem{theorem}{Theorem}
\newtheorem{lemma}{Lemma}
\newcommand{\caK}{\mathcal K}

\begin{document}

\preprint{APS/123-QED}

\title{Slow dissipation and spreading in disordered classical systems: \\ A direct comparison between numerics and mathematical bounds}

\author{Wojciech De Roeck}
\affiliation{K.U.Leuven University, Leuven 3000, Belgium}
\author{Francois Huveneers}
\affiliation{Department of Mathematics, King’s College London, Strand, London WC2R 2LS, United Kingdom}
\author{Oskar A. Prośniak}
\affiliation{Department of Physics and Materials Science, University of Luxembourg, L-1511 Luxembourg, G.\ D.\ Luxembourg}

\date{\today}

\begin{abstract}
We study the breakdown of Anderson localization in the one-dimensional nonlinear Klein-Gordon chain, a prototypical example of a disordered classical many-body system.
A series of numerical works indicate that an initially localized wave packet spreads polynomially in time, while analytical studies rather suggest a much slower spreading. 
Here, we focus on the decorrelation time in equilibrium. 
On the one hand, we provide a mathematical theorem establishing that this time is larger than any inverse power law in the effective anharmonicity parameter $\lambda$, 
and on the other hand our numerics show that it follows a power law for a broad range of values of $\lambda$.
This numerical behavior is fully consistent with the power law observed numerically in spreading experiments, and we conclude that the state-of-the-art numerics {may well be} unable to capture the long-time behavior of such classical disordered systems. 

\end{abstract}

\maketitle

\section{Introduction}
Since at least two decades there has been vivid interest in the dynamical behavior of certain disordered classical interacting systems, like the nonlinear discrete Schr\"odinger equation (NLS) or the nonlinear Klein-Gordon chain (KG).  
These systems are characterized by the feature that, when the anharmonicity is set to zero but the disorder strength remains finite, their eigenmodes are spatially localized by Anderson localization \cite{anderson_absence_1958}.  
The natural question is then what happens at small but finite anharmonicity, 
where there is an evident competition between the localization of the linear system on the one hand, and chaoticity and dissipation brought about by anharmonicity on the other hand. 

We are in particular motivated by previous compelling numerical work on spreading of initially localized wave packets in such systems \cite{pikovsky_destruction_2008,garcia-mata_delocalization_2009,flach_universal_2009,skokos_delocalization_2009,skokos_spreading_2010,mulansky_spreading_2010,terao_localization-delocalization_2011,larcher_subdiffusion_2012,min_subdiffusive_2012,yu_enhancement_2014,yusipov_quantum_2017,senyange_characteristics_2018,vakulchyk_wave_2019,kati_density_2020,skokos_frequency_2022}.
In these works, it was found that the width $w$ of the evolved wave packet grows as $w \varpropto t^{1/6}$ for several chains with a quartic on-site anharmonicity, including KG and NLS.    
These numerical findings are however at odds with a vast body of theoretical work \cite{frohlich_localization_1986,benettin_nekhoroshev-type_1988,poschel_small_1990,bourgain_wang_2007,johansson_kam_2010,wang_long_2009,fishman_perturbation_2009,fishman_nonlinear_2012,basko_weak_2011,cong_long-time_2021,de_roeck_asymptotic_2015,huveneers_drastic_2013}, supported by some numerics as \cite{oganesyan_pal_huse_2009,kumar_transport_2020}.  These works suggest that dissipative effects are non-perturbatively small in the effective anharmonicity $\lambda$, to be defined below. In particular, on the basis of  Hamiltonian perturbation theory similar to that used in the proof of the KAM theorem, it is predicted  that the thermal conductivity (or, for that matter, any other measure of dissipation) vanishes faster than any power of $\lambda$, which suggests a much slower spreading of the wavepacket, namely $w\varpropto (\log t)^{\alpha}$ for some power $\alpha$.

In the light of this discrepancy, the few pre-existing mathematical results like \cite{bourgain_wang_2007,wang_long_2009, huveneers_drastic_2013,cong_long-time_2021} do not offer solace. 
{Indeed, first, they do not apply to the system at hand, since the Anderson insulator at $\lambda=0$ is replaced in these works by a set of independent oscillators, known as the atomic limit of the system. 
And second, with the exception of \cite{bourgain_wang_2007}, they do not constrain the long-time limit of spreading of wave packets.}

\begin{figure}[ht]
\includegraphics[width=\linewidth]{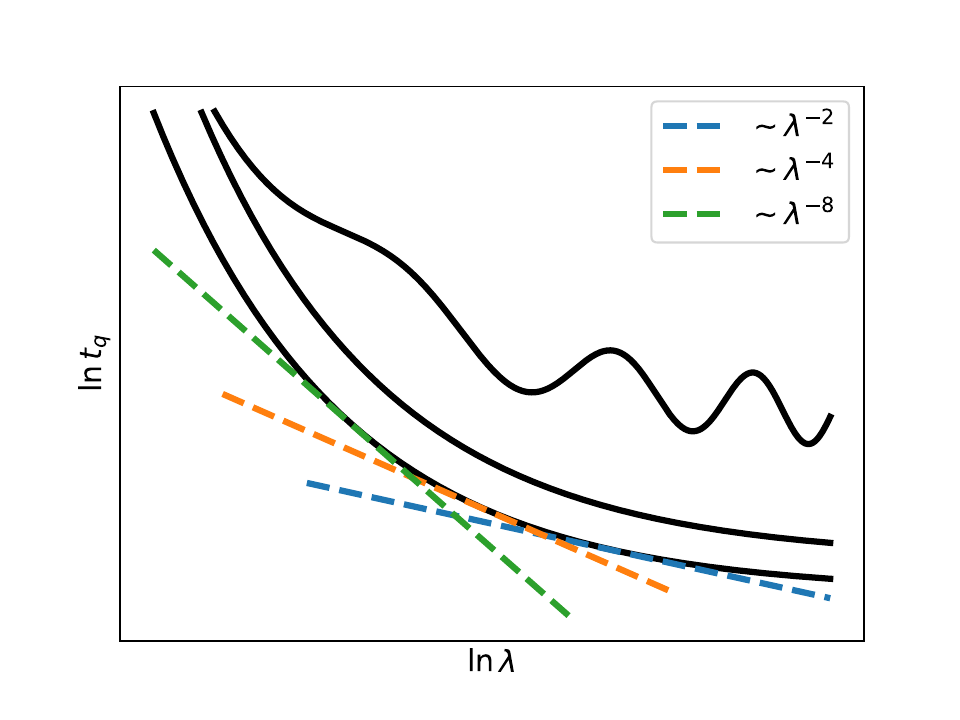}
\caption{\label{fig: gedankenplot} Mathematical constraints on the decorrelation time are shown as straight lines on this plot. The true curve has to lie above all these lines (like the 3 displayed black curves). 
}
\end{figure}

In this letter we report on work that, in our view, solves this puzzle: 
We provide for the first time a direct comparison between numerical observations and mathematical bounds for the KG chain. As an alternative to the speed of spreading or the diffusion constant, we focus on the decorrelation time in thermal equilibrium as it is more accessible to a proper mathematical treatment and to numerics. 
On the one hand, we prove a mathematical theorem stating that this time is larger than any polynomial in $\lambda$ as $\lambda \to 0$, see Figure \ref{fig: gedankenplot}.
On the other hand, we perform numerical analysis, for precisely the same system and the same observable, 
and we find a broad range of values of $\lambda$ where the decorrelation time scales as $\lambda^{-4}$, which is consistent with the diffusion constant scaling as  $\lambda^{4}$ {and with the behavior $w \varpropto t^{1/6}$ for the evolution of the width of a wave packet}. 

We conclude from our results that the KG chain is \emph{asymptotically many-body localized} as $\lambda \to 0$. By this term, we mean that the onset of dissipation is a fully non-perturbative phenomenon, {see} \cite{basko_weak_2011,huveneers_drastic_2013,De_Roeck_2014,de_roeck_asymptotic_2015,bols2018asymptotic} {as well as \cite{PhysRevLett.128.134102,zhang2023thermalization}}, implying in particular that decorrelation times grow faster than any power law in $\lambda$. 
Our findings also strongly suggest that the spreading of an initially localized wavepacket is slower than any polynomial in time, {and that the simulation time should be considerably extended in order to observe a behavior that deviate significantly from the scaling $w \varpropto t^{1/6}$}.

\section{Decorrelation times}    
The nonlinear Klein-Gordon Hamiltonian is  
$$
H=\sum_{i=1}^{L}\frac{p_i^2}{2}+\frac{\omega^2_i q_i^2}{2} +  \frac{ (q_{i+1}-q_i)^2}{2W} +\gamma \frac{q_i^4}{4},
$$
where the momenta $p_i$ are canonically conjugated to the positions $q_i$,
where $\omega^2_i$ are independent random variables, uniformly distributed in the interval $[1/2,3/2]$, 
where $W$ quantifies the disorder strength
and where we assume periodic boundary conditions $q_{L+1}=q_1$. 
We choose $W=4$ to align with \cite{flach_universal_2009,skokos_delocalization_2009}.
At $\gamma=0$, this Hamiltonian is quadratic and it can be recast in action-angle coordinates $(I_k,\phi_k)$ as 
\begin{equation}\label{eq: harmonic part}
  H_0=  \sum_k \nu_k I_k =\tfrac{1}{2}\sum_k  (\langle\psi_k,p\rangle^2+ \nu_k^2 \langle\psi_k,q\rangle^2 ).  
\end{equation}
Here $\langle \cdot, \cdot\rangle$ denotes 
{the usual scalar product in $\mathbb R^L$} and  $\nu_k^2,\psi_k$ are eigenvalues and eigenfunctions (normalized as $\langle \psi_k,\psi_k\rangle=1$) of the associated discrete Schrödinger operator $\mathcal H$:
\begin{equation}\label{eq: associated schrodinger}
    \mathcal H f(i)=\frac{1}{W}(f(i+1)+f(i-1)-2f(i))+ \omega_i^2f(i).
\end{equation}
Since this Schrödinger operator has a disordered potential and the spatial dimension is $1$, its eigenstates are Anderson localized \cite{anderson_absence_1958,gol1977pure,cmp/1103908590,10.1215/S0012-7094-82-04913-4}, i.e.\ they are exponentially localized {in space} around localization centers. As a consequence, the corresponding actions $I_k$ are exponentially localized as well.  An initially localized wavepacket will hence remain localized at all times, i.e.\  the width $w(t)$ is bounded above as $t\to\infty$, almost surely with respect to the disorder. 
At nonzero $\gamma$, $H$ contains also a quartic part in {square root of} the actions $I_j$, dependent on the angles, so that $I_j$ are no longer conserved.  
For a {generic} observable $X$, we introduce the decorrelation indicator
\begin{equation}\label{eq: def eta(t)}
    \eta_X(t)
    =
    \frac{1}{2} \frac{\langle (X(t)-X(0))^2\rangle}  {\langle X(t)^2\rangle-\langle X(t)\rangle^2}.
\end{equation}
with $\langle\cdot\rangle$ the thermal average at some fixed temperature $T$.
A scaling argument shows that $\eta_X(\cdot)$ depends on the parameters $(\gamma, T)$ only via the effective anharmonicity 
$$
\lambda=\gamma T.
$$
If the dynamics is thermalizing or more precisely \emph{mixing}, then we expect that $\eta_X(t)$ rises from $0$ to $1$ as $t\to\infty$, up to small corrections due to the difference between the canonical and microcanonical ensembles, vanishing as $1/L$ as $L\to\infty$.

{In our numerics, we choose $X=X_k=\nu_k I_k$ for a system of size $L=80$}, and we average over $L$ modes $k$ and over disorder, i.e.\@ over the variables $\omega_i^2$ (we use 200 realizations), obtaining the quantity 
$$
\eta(t)=\overline{\frac{1}{L}\sum_{k=1}^{L}\eta_{X_k}(t)}^{\,\omega}.
$$
We have run the dynamics up to a time $t=10^{10}$ for the smallest values of $\lambda$.
Figure \ref{fig: decorrelation profiles} shows our best estimator for $\eta(t)$  for several values of $\lambda$. 
{See Appendix~\ref{Appendix: Figure decorrelation} for further informations about our numerical procedure and Appendix~\ref{sec: quality numerics} for the robustness of our results.}
\begin{figure}[h]
 \includegraphics[scale=0.5]{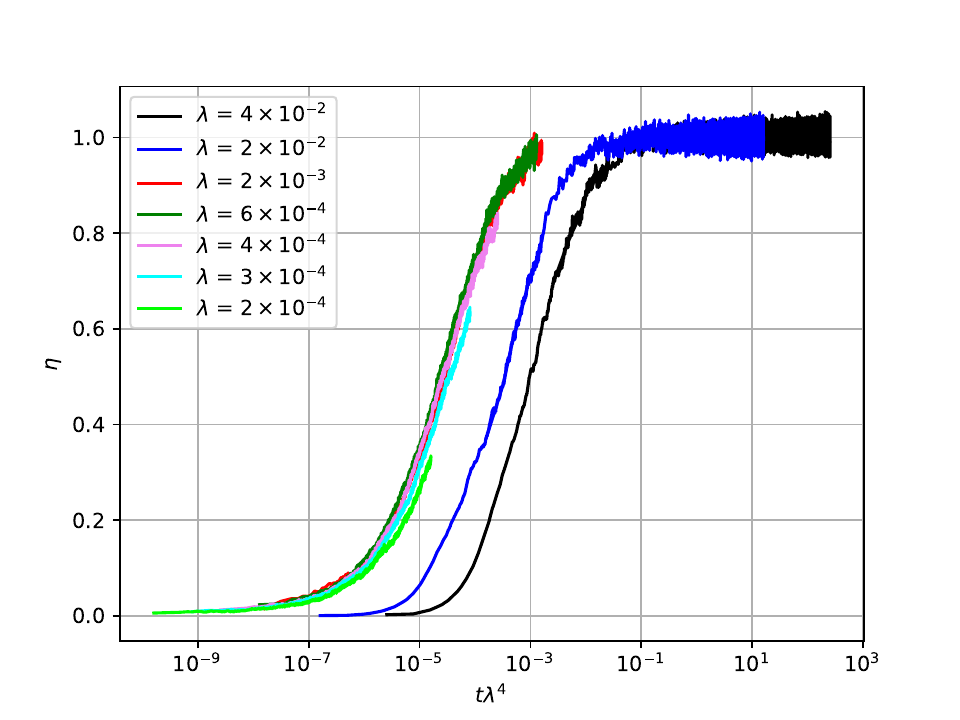}
\caption{\label{fig: decorrelation profiles} The decorrelation $\eta(t)$ as a function of rescaled time $\lambda^4 t$. With this rescaling, we observe collapse of decorrelation profiles for small anharmonicities $\lambda$.}
\end{figure}

\begin{figure}[h]
\includegraphics[scale=0.5]{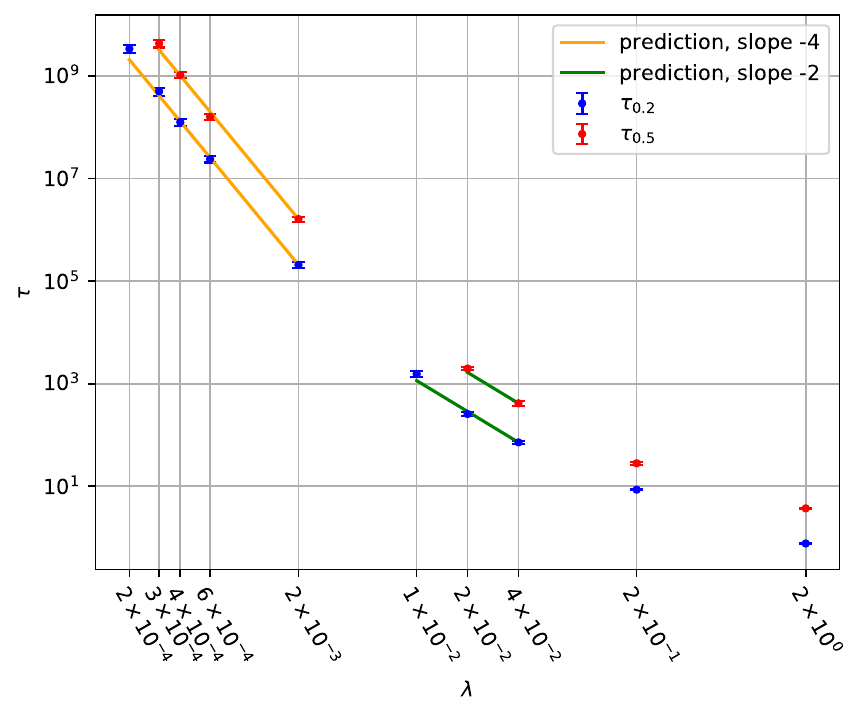}
 \caption{\label{fig: decorrelation time} The decorrelation time extraced from Figure \ref{fig: decorrelation profiles}, with $\eta_0=0.2$ and $\eta_0=0.5$.  We indicated by straight lines the regime for which $\tau\propto \lambda^{-4}$, and a short regime for which $\tau\propto \lambda^{-2}$.  }
\end{figure}

We define the decorrelation time $\tau$ as the time at which $\eta$ reaches a fixed value $\eta_0$ {(for clarity, we also use the notation $\tau_{\eta_0}$ when needed)}. 
We plot it for two different values of $\eta_0$ in Figure \ref{fig: decorrelation time}. 
Both figures \ref{fig: decorrelation profiles} and \ref{fig: decorrelation time} show the existence of scaling $\tau \propto \lambda^{-4} $ for small $\lambda$.

\section{Range of validity for scaling}
Both figures \ref{fig: decorrelation profiles} and  \ref{fig: decorrelation time} give a slight hint that the decorrelation time starts increasing faster than $\lambda^{-4}$ at small $\lambda$ but they also confirm that the $\lambda^{-4}$ scaling is valid for a range of anharmonicities that spans at least 1 or even 1,5 decades.  We can compare the data from Figures \ref{fig: decorrelation profiles} and \ref{fig: decorrelation time} to the numerics on spreading wavepackets by assuming that local thermal equilibrium holds inside the packet (we comment on this later): When the packet has a width $w$, the typical temperature inside the packet is of order $E_0/w$ with $E_0$ the total energy of the packet, 
see Appendix~\ref{sec: extracting anhar} for precise calculations. 
If we  \emph{assume} the scaling $w\propto t^{1/6}$ to hold for times in the range $[t_0,t_1]$, then that scaling holds for a corresponding range of temperatures $[T_0,T_1]$ with 
$T_1/T_0 = (t_1/t_0)^{1/6}$. Looked upon differently, this means that the validity of the scaling regime for $1,5$ decades correspond to its validity for $9$ decades in time!
Indeed, in \cite{flach_universal_2009,skokos_delocalization_2009}, the scaling $w\propto t^{1/6}$ was observed for $6-8$ decades, which is hence perfectly consistent with out results.    
In \cite{vakulchyk_wave_2019}, it was observed for 12 decades in a related, though different, model. 
While a direct comparison with our numerics is difficult due to the differences between the models, it is important to realize that here as well, the relevant parameter is the effective anharmonicity reached when the packet has maximally spread out, see 
Appendix~\ref{sec: flach comparison} for details.
Note that the above reasoning does not rely on any intrinsic connection between the $w\propto t^{1/6}$ law and the $\tau \propto \lambda^{-4}$ law. 
Indeed, although we will explain later that \emph{there is such a connection}, this is actually not crucial for our argument.

Finally, we can compare our numerics with the numerics on spreading of wave packets in the following way. Since we see a hint of a deviation from the $\tau\propto \lambda^{-4}$ law, one could ask how hard it would be to probe the same nonlinearities as we do in a spreading experiments. It is harder because spreading numerics needs a larger volume than our decorrelation numerics. More concretely, to see our smallest nonlinearity in the setup \cite{flach_universal_2009,skokos_delocalization_2009}, we would have to simulate the system at least $10^2$ times longer than done in \cite{flach_universal_2009,skokos_delocalization_2009}, (assuming that the spreading does not slow down compared to the $t^{1/6}$ law), see Appendix~\ref{sec: going smaller}.

\section{A rigorous result} 
In the companion paper~\cite{de_roeck_notitle_2023}, we have proven the following mathematical result. 
Let us fix an arbitrary threshold value $0<q<1$ for the decorrelation parameter $\eta$, as in Figure \ref{fig: decorrelation time}.  
{We remind that} $\tau_q$ is the smallest  positive time such that $\eta(t)\geq q$ (this time exists by continuity). Then

\begin{theorem}[Slow decorrelation]\label{thm: decorrelation time}
For any integer $n$, there is a constant $c_n>0$ such that
$$
\liminf_{L\to\infty} \,  \tau_q  \geq c_n  |\lambda|^{-n} \qquad \text{with probability } 1.
$$  
\end{theorem}
Hence, this theorem clearly states that the curve in Figure \ref{fig: decorrelation time} has to curve upwards eventually, as shown in \ref{fig: gedankenplot}.  
The combination of the above theorem and the numerics shown above, teaches us that the numerics is not able to capture the genuine asymptotic behaviour at small $\lambda$. Indeed, the numerics suggests a $\lambda^{-4}$ law, whereas we rigorously know that it cannot be more than a transient behaviour. As was already argued above, the smallest effective anharmonicity reached in our numerics for the decorrelation time, is of the same order as the effective anharmonicity reached in the spreading numerics, so it seems reasonable to posit that also the latter is not yet probing the true asymptotic regime. 

A full mathematical proof of Theorem~\ref{thm: decorrelation time} is provided in our companion paper~\cite{de_roeck_notitle_2023}, and here we only highlight the main ideas.  
See also Appendix~\ref{sec: proof of theorem} for a detailed description of the proof. 

For simplicity, we set $T=1$, so that $\gamma$ is identified with $\lambda$. 
We write the KG Hamiltonian as $H=H_0+\lambda H_{1}$, where $H_0$ is the harmonic part, and we let $n$ be the integer featuring in Theorem~\ref{thm: decorrelation time}. 
For $X = X_k = \nu_k I_k$, the strategy is to cast its time derivative $dX/dt = \{ H,X\}$ as
\begin{equation}\label{eq: commutator equation}
   \{H,X\}=\lambda \{H,u_n\} +\lambda^n  g_n ,
\end{equation}
where $\{\cdot,\cdot\}$ denotes the Poisson bracket. 
Here $u_n,g_n$ are well-bounded with large probability with respect to disorder, more precisely,
$$
\mathrm{Prob} (\max(\langle u_n^2\rangle, \langle g_n^2\rangle )\geq M) \leq M^{-s}
$$
for some small but finite $s>0$, and for any $M>0$. 
The representation~\eqref{eq: commutator equation} takes the integral form
$$
X(t)-X(0)= \lambda  (u_n(t)- u_n(0)) + \lambda^n \int_0^t ds\,  g_n(s),
$$
which, applying the Cauchy-Schwartz inequality a few times, yields
\begin{align*}
 \langle (X(t)-X(0))^2 \rangle & \leq 4\lambda^2(\langle u^2_n(t) \rangle +\langle u^2_n(0)\rangle) \\[1mm]
 &  \qquad +  2\lambda^{2n}t^2 \sup_{0\leq s \leq t}
\langle g^2_n(s) \rangle.  
\end{align*}
Using the invariance of the thermal average $\langle \cdot\rangle$ under time evolution, and the probabilistic bound stated above, 
we get hence that  $\langle (X(t)-X(0))^2 \rangle \leq 2M(4\lambda^2+\lambda^{2n}t^2)$ with probability at least $1-M^{-s}$. 
We have hence shown that for a fixed mode $k$, $X(t)-X(0)$ remains appropriately small for a long time, and with large probability. 
By scaling $M$ with $\lambda$ and with system size, and dealing with correlations between different modes $k,k'$, one then derives Theorem \ref{thm: decorrelation time}.

The crucial equation \eqref{eq: commutator equation} is derived via perturbation theory. 
Given a polynomial $f$ in $p,q$ there exists a well-bounded observable $h$ such that 
\begin{equation}\label{eq: local inverse}
    \{H_1,f\}= \{H_0,h\},
\end{equation}
provided that $\{H_1,f\}$ is $p$-antisymmetric. 
This applies to our set-up since $\{H,X\}=\lambda\{H_1,X\}$ is indeed $p$-antisymetric, and an iterative use of eq.~\eqref{eq: local inverse} yields the representation~\eqref{eq: commutator equation}.

In solving eq.~\eqref{eq: local inverse} by inverting the Liouville operator $\{H_0,\cdot\}$, one encounters random expressions of the form
\begin{equation}\label{eq: z like expression}
Z(i)= \sum_K \frac{\prod_{k\in K}  |\psi_k(i)| }{\sum_{k\in K} \sigma_k\nu_k},
\end{equation}
for a fixed site $i$, where $K$ ranges over collections of distinct $n$-tuples of modes $k$, and $\sigma_k =\pm 1$. Such expressions are familiar from KAM techniques, with divergent denominators corresponding physically to resonances.

The main technical point in our proof is the control of \eqref{eq: z like expression}. 
In particular, we prove that
$\mathrm{Prob}(Z(i)>M) \leq M^{-s}$ for some power $s>0$ to derive the statements below \eqref{eq: commutator equation}. 
To prove this bound, we need to control correlations (see e.g.\ Lemma~\ref{lem: bound on delta in paper} in Appendix~\ref{sec: proof of theorem}) between different eigenvalues $\nu^2_k$, namely those that appear in the denominator of \eqref{eq: z like expression}. At the time of writing, results on such correlations are not available in the mathematical literature, except in the case where the eigenvalues are close to each other (this goes under the name Minami bound, see \cite{minami1996local,graf2007remark,aizenman2008joint}), or in the case of just two eigenvalues, see \cite{klopp2011decorrelation}. 
Details on this problem are provided in the companion paper \cite{de_roeck_notitle_2023}. Actually, the challenge of dealing with eigenvalue correlations  prohibits us from extending our theorem to, for example, the discrete nonlinear Schrödinger equation with disorder. Indeed, in such a system, KAM-like perturbation theory leads to  \eqref{eq: z like expression} but with $\sum_k \sigma_k\nu_k$ replaced by $\sum_k \sigma_k\nu^2_k$, see \cite{fishman_perturbation_2009,fishman_nonlinear_2012}.

\section{Connection between scaling laws}  
While this is not central, it is certainly desirable to connect
the apparent $\lambda^{-4}$ law for the decorrelation time with the (apparent) $t^{1/6}$ law for wave packet spreading. {We now turn to this.} Let us assume that the spreading is governed by a nonlinear diffusion equation for the energy density $\rho(x,t)$, i.e.\  $\partial_t\rho(x,t)=-\partial_x D(\rho) \partial_x \rho(x,t)$. 
A computation \cite{tuck_explicit_1976}  shows then that the $t^{1/6}$-spreading law corresponds to the scaling behaviour $D(\rho)\propto \rho^4$ for the diffusivity $D$. Since $\rho\propto T$, this is equivalent to $D(\rho)\propto \lambda^4$. We now connect this scaling of $D$ with the time-decorrelation function $\eta(\cdot)$. We assume that the hydrodynamic mode, i.e.\ transport of energy, is the slowest contributing process, and hence the one that determines $\eta(t)$ at long times. Within a fluctuating hydrodynamics model defined on a mesoscopic scale, where in particular the disorder is no longer visible, we can write $\partial_t (\delta \rho) =\partial_x (D\partial_x (\delta \rho)+a\zeta) $  with $\delta\rho$ small deviations from the spatially uniform density, $\zeta$ space-time white noise $\langle \zeta(x',t') \zeta(x,t)\rangle=\delta(x'-x)\delta(t'-t) $, and $D$ and $a$ related by the fluctuation-dissipation theorem. For more details and computations, we refer to Appendix~\ref{sec: linear hydro}. In particular, within this model, one computes $\eta(t) \approx 1-\frac{1}{\sqrt{Dt}} $, and hence $D(\rho)\propto \lambda^4$ is equivalent to the collapse seen in Fig.\ \ref{fig: decorrelation profiles} at small $\lambda$ and to the line with slope $-4$ in Fig.\ \ref{fig: decorrelation time}.

\begin{figure}[ht]
\centering
\includegraphics[scale=0.6]{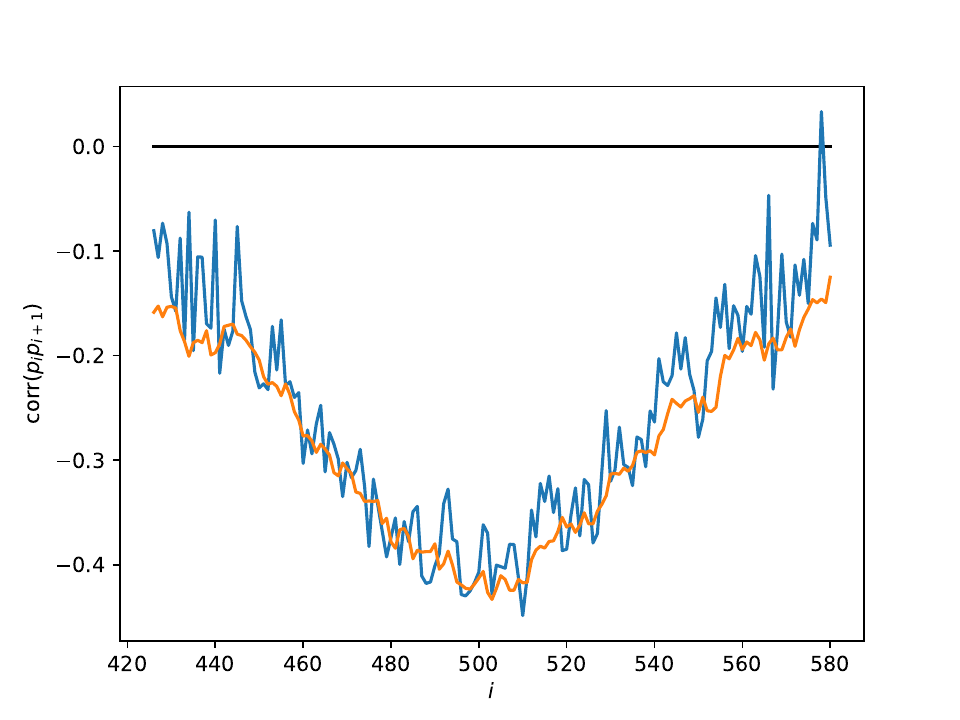}
\caption{\label{fig: correlation} Correlation $\text{corr}(p_i,p_{i+1})$. The blue curve is a numerical estimate for the ergodic average of $\text{corr}(p_i,p_{i+1})$ after evolving the system of length $L=10^3$ from the same initial conditions as considered in \cite{flach_universal_2009} for $t=10^8$.
The orange curve shows  $\text{corr}(p_i,p_{i+1})$  in the prethermal ensemble constructed from data in a window of size $10$ around each site.  
The black line is the value of $\text{corr}(p_i,p_{i+1})$ in a thermal state. 
See Appendix~\ref{sec: details for figure pretherm} for further explanations.}
\end{figure}

Of course, the above reasoning presupposes that 
the wave packet is locally well-described by an equilibrium state, with temperature varying throughout the packet.   
This is actually far from obvious for a system with a finite amount of energy. 
Yet, a close inspection of correlations in the wave packet reveals that a description by a statistical ensemble may well be accurate, 
though the packet is not properly in equilibrium. 
Indeed, as {a representative} example, we show in figure~\ref{fig: correlation} that the correlation between $p_i$ and $p_{i+1}$ is correctly captured by a pre-thermal ensemble characterized by two (pseudo-)conserved quantities: energy and the total action $I = \sum_k I_k$ of the harmonic system \eqref{eq: harmonic part}. 
The latter quantity is only approximately conserved, so that this ensemble is pre-thermal rather than properly thermal, see \cite{mendl_lu_lukkarinen_2016,PhysRevResearch.2.022034} for an extensive analysis and discussion of this prethermal phase in the disorder-free Klein-Gordon chain. For the sake of justifying transport equations, this prethermal phase serves just as well as the thermal phase.

\section{Conclusion}
We investigated the disordered Klein-Gordon chain with an anharmonic perturbation.  A mathematical theorem shows that the decorrelation time $\tau$ for this chain grows faster than any polynomial in $1/\lambda$: the chain is \emph{asymptotically localized}. On the other hand, numerics is consistent with the scaling $\tau\propto 1/\lambda^4$ for a broad range of values. The latter scaling corresponds to the anomalous spreading of wavepackets $w(t)\propto t^{1/6}$ observed in a series of previous works. 
As far as we can ascertain, state-of-the-art numerical experiments reproduce reliably the dynamics of the chain, but fall short of capturing the true long-time behaviour of the system. We conclude that the anomalous spreading law is most likely a transient behaviour, giving way at later times to a spreading law smaller than any polynomial.

\bibliographystyle{apsrev4-2}
 
\let\oldaddcontentsline\addcontentsline
\renewcommand{\addcontentsline}[3]{}
\bibliography{KGbiblio}
\let\addcontentsline\oldaddcontentsline

\pagebreak
\phantom{a}
\pagebreak

\appendix
\section{About the proof of Theorem~\ref{thm: decorrelation time}}\label{sec: proof of theorem}

We provide a detailed description of the proof of Theorem~\ref{thm: decorrelation time}. 
We refer to our companion paper \cite{de_roeck_notitle_2023} for a mathematically rigorous proof. 
 
\subsection{Creation and annihilation functions}
Recall that we write $\psi_k$ and $\nu^2_k$ for the normalized eigenvectors and (positive) eigenvalues of the discrete Schrodinger operator $\mathcal H$ defined in \eqref{eq: associated schrodinger}.
Let us define creation $a_k^+$ and annihilation $a_k^-$ functions by
$$
a^{\mp}_k=\frac{1}{2} \sum_j \psi_k(j) \left( {\sqrt{\nu_k}}  q_j \pm   
\frac{ \mathrm{i} p_j}{\sqrt{\nu_k}}  
\right).
$$
These functions satisfy the familiar canonical commutation relations
$$
 \{a^{+}_k, a^{-}_{k'}  \}= \mathrm{i} \delta_{k,k'}  \qquad \{a^{+}_k, a^{+}_{k'}  \} =\{a^{-}_k, a^{-}_{k'}  \}=0 .
$$
The harmonic part $H_{\mathrm{har}}$ of the Hamiltonian is then expressed as
$$
H_{\mathrm{har}}= \sum_k \nu_k a^+_k a^-_k .
$$
We introduce the following abbreviation. We write $\caK$ for the set of pairs $K=(k,\sigma)$ with $k$ labeling the modes (as above) and $\sigma \in\{-1,1\}$.
We associate to any $K$ a creator/annihilator function by 
$$a_K=a_k^\sigma,\qquad K=(k,\sigma).$$
In practice, we will use powers $\caK^n$ of $\caK$ and we write in general $\underline{K}=(K_1,\ldots, K_n)$ for elements in $\caK^n$. 
In particular, we abbreviate
$$
a_{\underline{K}}=a_{K_1}\ldots a_{K_n}.
$$
We can now handily represent the anharmonic part of the Hamiltonian
$$
H_{\mathrm{an}}= \sum_{\underline{K}\in \caK^4}   H_{\mathrm{an}}(\underline{K}) a_{\underline{K}}
$$
with
$$
H_{\mathrm{an}}((K_1,\ldots,K_4)) = \tfrac{1}{16}\sum_{j=1}^L \prod_{i=1}^4 \frac{\psi_{k_i}(j)}{\sqrt{\nu_{k_i}}},
$$
i.e.\ this expression is independent of the variables $\sigma_i$.\\

If $K=(k,\sigma)$, let us write $-K=(k,-\sigma)$, i.e\ the same mode but opposite sign.   
We say an element $\underline{K} \in \caK^{n}$ is \emph{paired} if its components $K_1,\ldots,K_n$ can be partitioned into pairs 
such that each pair is of the form $({K},-{K})$. In particular, if $n$ is odd, no element is paired.
For any $\underline{K} \in  \caK^{n}$, we introduce
$$
\Delta(\underline{K})=\sum_{i=1}^n \sigma_i k_i.
$$
Note that $\Delta(\underline{K})=0$ if  $\underline{K}$ is paired.  
If $\underline{K}$ is not paired, $\Delta(\underline{K})$ can of course still vanish accidentally, but this happens with probability zero. 
In fact, we prove in \cite{de_roeck_notitle_2023} the following quantitative estimate, for any $\epsilon>0$:
\begin{lemma}\label{lem: bound on delta in paper}
$$\mathbb{P}\left(\min_
{\substack{   \underline{K} \in \caK^m    \\ \underline{K} \text{not paired}  }} |\Delta(\underline{K})| \leq \epsilon \right) \leq  L^m \epsilon^{\frac{1}{m+1}}.
$$
\end{lemma}
The proof of this lemma is not short, but the lemma itself is very intuitive: the different eigenvalues $\nu^2_k$ of the disordered Schrodinger operator $\mathcal{H}$ can, for many purposes, be considered to be uncorrelated variables.

A central point of our reasoning is the possibility of inverting the harmonic Liouville operator $\{H_{\mathrm{har}},\cdot\}$ on local functions.  

\begin{lemma} \label{lem: inverse not paired}
If $\underline{K}$ is not paired, the equation
$$\{H_{\mathrm{har}}, u\}= a_{\underline{K}} $$
is solved by 
$$
u=   \frac{1}{\Delta(\underline{K})} a_{\underline{K}}
$$
\end{lemma}
Let us now consider a homogenous polynomial of degree $d$ in the functions $a_{K}$, i.e.\ 
$$
h=\sum_{\underline{K}\in \caK^d} \hat{h}(\underline{K})a_{\underline{K}},
$$
for some coefficients $\hat h(\underline K)$. 
One can check that homogenous polynomials of degree $d$ are also homogenous of degree $d$ when written as functions of the original $p,q$-variables.
It will be of particular interest for us to consider functions $h$ that are anti-symmetric under the transformation $$
(K_1,\ldots,K_n)\to (-K_1,\ldots,-K_n), $$
i.e.\ under flipping all $\sigma_i$. In the orignal variables, this corresponds to functions that are antisymmetric under time-reversal, i.e.\ under flipping all momenta. 
The following observation is then easily checked.
\begin{lemma}\label{lem: anitsymm not paired}
Let
$$
h=\sum_{\underline{K}\in \caK^n}   \hat{h}(\underline{K}) a_{\underline{K}}
$$
be  antisymmetric under time-reversal.
Then, 
$$
h=\sum_{\substack{\underline{K}\in \caK^n \\  \underline K \text{not paired}}}   \hat{h}(\underline{K}) a_{\underline{K}},
$$
i.e.\ one can set the coefficient of paired $\underline K$ to zero.
\end{lemma}

\subsection{The recursive scheme}

Let us call $f^{(1)}=\{H,X_{k_0}\}$ for some fixed choice of mode $k_0$ and with $X_{k_0}=\nu_{k_0}a^+_{k_0}a^-_{k_0}$ the energy of mode $k_0$.
We will now derive the representation claimed in the main text
\begin{equation}\label{eq: repetition fone rep}
f^{(1)}= \lambda \{H,u_n\} +\lambda^n g_n   
\end{equation}
with explicit expressions for $u_n,g_n$. 
By inspection, we see that $f^{(1)}$ is anti-symmetric under time reversal. Therefore, by Lemmas \ref{lem: inverse not paired} and \ref{lem: anitsymm not paired}, we can find a function $u^{(1)}$ that solves the equation 
$\{u^{(1)},H_{\mathrm{har}}\}=f^{(1)}$. 
Defining now $f^{(2)}=-\{u^{(1)},H_{\mathrm{an}}\}$, we  check that $f^{(2)}$ is again antisymmetric, since $H_{\mathrm{an}}$ is symmetric.  The strategy is to iterate this structure, defining in general
$$
f^{(i)}=-\{H_{\mathrm{har}}, u^{(i)}\}, \qquad f^{(i+1)}= -\{u^{(i)},H_{\mathrm{an}}\}.
$$
We use the first equation to determine $u^{(i)}$ from $f^{(i)}$ (using again Lemma's \ref{lem: inverse not paired} and \ref{lem: anitsymm not paired}) and the second equation to determine $f^{(i+1)}$ from $u^{(i)}$.  These definitions hence yield, using $H=H_{\mathrm{har}}+\lambda H_{\mathrm{an}} $,
\begin{align}
  f^{(i)} = \{u^{(i)},H_{\mathrm{har}} \} &=   \{ u^{(i)},H\}
-\lambda  \{ u^{(i)},H_{\mathrm{an}}\} \\
&=\{ u^{(i)},H\}
+\lambda f^{(i+1)}.   
\end{align}
Iterating this relation we arrive at  \eqref{eq: repetition fone rep} with 
$$
u_n=\sum_{i=1}^n \lambda^{i-1}u^{(i)}, \qquad g_n=f^{(n+1)}.
$$

We are now fully equipped to construct explicitly these functions and to investigate their properties. We focus on the function $f^{(n)}$.

\subsection{Explicit expression for $f^{(n)}$}

We find $f^{(n)}$ as a sum over $\underline{K}\in \caK^{4n}$.  For $\underline{K}\in \caK^{4n}$ we write $\underline{K}=(\underline{K}^{(1)},\ldots, \underline{K}^{(n)})$ where each $\underline{K}^{(j)} \in \caK^4$, and we also write 
$\underline{M}^{(j)}=(\underline{K}^{(1)},\ldots, \underline{K}^{(j)}) \in \caK^{4j}$.  We set $\chi(\underline{K})=0$ when at least one of the $\underline{M}^{j}$ is paired, and $\chi(\underline{K})=1$ otherwise.

Then, following the construction above, we get
\begin{align} \label{eq: explicit expression f}
  f^{(n)} &= \sum_{\underline{K} \in \caK^{4n}} \chi(\underline{K})
f^{(1)}(\underline{K}^{(1)})   H_{\mathrm{an}}(\underline{K}^{(2)})\ldots    H_{\mathrm{an}}(\underline{K}^{(n)})
  \\  
& \qquad   \qquad  \prod_{j=1}^n \frac{1}{\Delta(\underline{M}^{j})}  \quad \{ a_{\underline{K}^{(n)}},   \ldots     \{ a_{\underline{K}^{(2)}} , a_{\underline{K}^{(1)}}  \} \ldots
\}
\end{align}
To ease the analysis later on, we abbreviate this as 
\begin{align} \label{eq: explicit expression f repeat}
  f^{(n)} &= \sum_{\underline{K} \in \caK^{4n}}
\frac{\mathrm{coeff}(\underline{K} )}{D(\underline{K} )}   A(\underline{K} )
\end{align}
where $A(\underline{K} )$ stands for the nested Poisson brackets and $D(K)$ stands for the product of $\Delta(\cdot)$ factors. 

This expression was derived using two ideas:
\begin{enumerate}
    \item For every $f^{(i)}$ with $i\leq n$, we used Lemma \ref{lem: anitsymm not paired} to set the coefficients of paired $\underline{M}^{(j)}$ to zero.
    \item Given $\underline{K} \in \caK^m,\underline{K'} \in \caK^{m'}$, the bracket 
    $$
    \{a_{\underline{K}},a_{\underline{K}'}\} $$ is a sum of some $\pm a_{\underline{K}''}$ with $K''\in \caK^{m+m'-2}$. Each of the $\underline{K}''$ appearing in this sum satisfies 
    $$
    \Delta (\underline{K}'')= \Delta ((\underline{K},\underline{K}'))
    $$
    where $(\underline{K},\underline{K}')$ is simply a concatenation of the arrays $\underline{K},\underline{K}'$.
    This was used to keep the expression inside $\Delta(\cdot)$ simple. 
\end{enumerate}

\subsection{Analysis of $f^{(n)}$}

The function $f^{(n)}$ obtained above, is a homogeneous polynomial in the functions $a_K$ of degree $4n-2(n-1)=2n+2$. To obtain this, we took into account that each bracket $\{\cdot,\cdot\}$ reduces the degree by $2$. 
The number of monomials grows rapidly in $n$  and system size $L$, but Anderson localization of the modes $k$ leads to good bounds on $f^{(n)}$, as we explain now.  
To keep the discussion intuitive, it is handy to have a cartoon picture of Anderson localization. We therefore assume that all modes $k$ are assigned a localization center $x(k)$, and that they decay exponentially away from that center with a uniform decay rate $\kappa$, i.e.\ the corresponding eigenfunction $\psi_k$ is bounded as 
\begin{equation}\label{eq: simplified decay}
    |\psi_k(j)|\leq \frac{1}{C_{2\kappa}}e^{-\kappa |j-x(k)|}, \qquad \text{with}\quad  C_\alpha=\sum_{i\in \mathbb{Z}} e^{-\alpha |i|}
\end{equation}
The reason that this is a cartoon is because in reality, eigenfunctions can decay slower at finite distance. 
See \cite{de_roeck_notitle_2023} for a fully rigorous argument, devoted of any simplifying assumption. 

The first crucial observation about \eqref{eq: explicit expression f} is that all modes $K=(k,\sigma)$ that appear in the expression, are connected to $k_0$ in the following sense: Let $x$ be the localization center of the mode $k$ appearing in $\underline{K}$ that maximizes the distance 
$$\ell(\underline{K})=|x(k_0)-x(k)|,$$
then the coefficient $\mathrm{coeff}(K)$  in \eqref{eq: explicit expression f repeat} is bounded by 
\begin{equation}\label{eq: bound on coeff}
  |\mathrm{coeff}(\underline{K} )| \leq e^{-2\kappa \ell(\underline{K})} N_n,
  \quad  
  N_n=  \left(\frac{1}{C_{2\kappa}}\right)^{4n} \times \left(\frac{C_{\kappa}}{\nu_{\min}^4}\right)^n  .
\end{equation}
Here, the last factor $C_{\kappa}/\nu_{\min}^4$
is a crude estimate for the sum over lattice sites inside the expression for $H_{\mathrm{an}}(\underline{K} )$. 
We will see later on that the precise form of $N_n$ is irrelevant for our purposes, as long as $N_n$ is uniformly bounded as a function of system size $L$.

A second observation goes as follows. Let us fix $\ell$ as above, so that all modes actually fit into a box $B(k_0,\ell)$ of half-size $\ell$ centered at $x(k_0)$. Then the statistics of the denominator, which depends on the modes only through the eigenvalues $\nu^2_k$, can actually be studied in the this box $B(k_0,\ell)$. This means that we can apply Lemma \ref{lem: bound on delta in paper} with the system size $L$ replaced by the box size $|B(k_0,\ell) |=2\ell$, which will be crucial in our estimates.

Using these approximations, we now estimate \eqref{eq: explicit expression f repeat} by summing over the (introduced above) variable $\ell$
\begin{equation}
    \label{eq: simplified f to bound}
    |f^{(n)}| \leq  N_n  \sum_{\ell=0}^{\infty} e^{-2\kappa\ell}  \left(\sum_{\underline{K}\in \caK_\ell^{4n}}  |A(\underline{K})|\right) \left(  \max_{\underline{K} \in \caK_\ell^{4n} }  \frac{\chi(\underline{K})}{|D(\underline{K})|}  \right)
\end{equation}
where we have written $\caK_\ell$ to indicate that we only consider modes inside the box $B(k_0,\ell)$ and not in the full volume. 

\subsection{Proof of required bounds on $f^{(n)}$}

As indicated in the main text, we need to prove the following probabilistic bound:
\begin{lemma}\label{lem: proba bound of f}
There is a $s>0$ that can depend on $n$ but not on the system size $L$ such that 
\begin{equation}\label{eq: proba bound to obtain}
    \mathbb{P}(\langle |f^{(n)}|^2 \rangle \geq M) \leq M^{-s},  \qquad \text{for any $M>0$}
\end{equation}
where $\langle \cdot \rangle$ is the thermal expectation at unit temperature, and the probability $\mathbb{P}$ refers to disorder. 
\end{lemma}

Strictly speaking, in the main text we need this lemma for $g_n$ (which is equal to  $f^{(n+1)}$) and for $u_n$. However, the argument for $u_n$ is completely analogous to that for $f^{(n+1)}$, and we omit it here.

As above, we sketch the proof of this lemma under the simplifying assumptions outlined above, starting with \eqref{eq: simplified decay}.
Let us deal first with the thermal expectation value $\langle \cdot\rangle$. The only expression on the right hand side of  \eqref{eq: simplified f to bound} that is non-constant on phase space, is $|A(\underline K)|$. Due to the contractions occurring when evaluating the nested Poisson brackets, $A(\underline{K})$ is actually a sum of monomials $a_{\underline{K}'}$ with $\underline{K}' \in \caK^{2n+2}$.   
When taking the square in \eqref{eq: proba bound to obtain}, we will have hence to bound 
\begin{equation}\label{eq: biga}
    |\langle A_{\underline{K}'}A_{\underline{K}''}\rangle| \leq  \sqrt{|\langle A^2_{\underline{K}'}\rangle \langle  A^2_{\underline{K}''}\rangle| } 
\end{equation}
for $\underline{K}',\underline{K}'' \in \caK^{2n+2}_\ell$.
Such monomials in creation/annihilation functions have a finite $n$-dependent expectation value in a thermal state. 
The number of such monomials  in this sum is also  bounded by an $n$-dependent factor. Since the $n$-dependence is irrelevant for us here, we simply bound \eqref{eq: biga} by a generic function of $n$ that we call $N_n$, just like the one appearing in \eqref{eq: bound on coeff}. The symbol $N_n$ below can hence mean different functions of $n$, but always bounded uniformly in $L$.

Hence, we obtained the bound
\begin{equation}\label{eq: new bound simple}
\langle |f^{(n+1)}|^2 \rangle \leq N_n \sum_{\ell} e^{-2\kappa \ell}
R(\ell)
\end{equation}
where we have abbreviated 
$$ R_n(\ell)= \max_{\substack{   \underline{K} \in \caK_\ell^{4n}   }} \frac{\chi(\underline K)}{D(\underline K)^2}.
$$
For future use, we also introduce 
$$ r_j(\ell)= \max_{\substack{   \underline{M} \in \caK_\ell^{4j}    \\ \underline M \text{not paired}  }} \frac{1}{\Delta (\underline M)^2}
$$
such that 
\begin{equation}\label{eq: factor r}
R_n(\ell) \leq \prod_{j=1}^n r_j(\ell).
\end{equation}
Comparing \eqref{eq: new bound simple} with \eqref{eq: proba bound to obtain}, we see that it suffices to  prove
\begin{equation}\label{eq: bound on rell}
   \mathbb{P}(\sum_{\ell} e^{-2\kappa \ell}  R(\ell)   \geq M) \leq N_n M^{-s} .
\end{equation}

Indeed, by adjusting $s$, the lemma will follow from the bound \eqref{eq: bound on rell}.  We now hence focus on proving \eqref{eq: bound on rell}.
Using first a Markov inequality, and then subadditivity of the function $x\mapsto x^s$, this probability is bounded by 
\begin{align}
   & \mathbb{P}\left(\sum_{\ell} e^{-2\kappa \ell}  R(\ell)   \geq M \right)  \\
   &\qquad \leq 
M^{-s}\mathbb{E}\left((\sum_{\ell} e^{-2\kappa  \ell}     R(\ell)  )^s\right) \\
& \qquad \leq    M^{-s} \sum_{\ell} e^{-2s\kappa  \ell} \mathbb{E}\left((R(\ell))^s\right).  \label{eq: penultimate bound pr}
 \end{align}
 To continue, we use \eqref{eq: factor r}:
 \begin{align}
    &  M^{-s} \sum_{\ell} e^{-2s\kappa  \ell} \mathbb{E}\left((R(\ell))^s\right)  \\
&\leq    M^{-s} \sum_{\ell} e^{-2s\kappa  \ell} \mathbb{E}\left(\prod_{j=1}^n r_j(\ell)^s\right) \\
& \leq  M^{-s}  \sum_{\ell}   e^{-2s\kappa  \ell}\prod_{j=1}^n    \mathbb{E}\left(( r_j(\ell)^{s n_*})\right)^{\frac{1}{n_*}}\label{eq: last bound pr}
\end{align}
where $n_*=\lceil\log_{2}(n)\rceil$. 
To get the last inequality, we applied the Cauchy-Schwarz inequality $n_*$  times.
To estimate the expectation of powers of $r_j(\ell)$, we consider the dyadic decomposition
$$
\mathbb{E}((r_j(\ell))^{n_*s} )\leq \sum_{i=0}^\infty \mathbb{P}(  2^{i}-1 \leq r_j(\ell) \leq 2^{i+1}-1   )  2^{(i+1) n_* s}   
$$
and we apply Lemma \ref{lem: bound on delta in paper} for each $j$ and $i$, yielding 
$$
\mathbb{P}(  2^{i}-1 \leq r_j(\ell)   ) \leq \ell^{j} (2^{i}-1)^{-\frac{1}{2+2j}}
$$
We can now choose $s$ small enough, in particular satisfying
$$
n_*s-\frac{1}{2+2n} <0,
$$
such that the sum over dyadic scales $i$ is convergent. We then obtain 
$$
\max_{j=1}^n\mathbb{E}((r_j(\ell))^{n_*s} )\leq \ell^n N_n
$$
Plugging these bounds into \eqref{eq: last bound pr} and \eqref{eq: penultimate bound pr}, we hence obtain \eqref{eq: bound on rell}. As explained earlier, this implies \eqref{eq: proba bound to obtain} and therefore we have finished the sketch of proof of Lemma \ref{lem: proba bound of f}.

\section{Linear fluctuating hydrodynamics}\label{sec: linear hydro}
We provide some details on linear fluctuating hydrodynamics, which is used in the main text to derive the normal diffusive behaviour for the decorrelation measure $\eta(t)$.
In the definition of $\eta(t)$, we replace the observables $X_k$ by the hydrodynamic local fields $\phi_x$, i.e.\ the local energy density.

\subsection{Setting}
We rename the field as $\phi_x(t)=\delta\rho_x(t)$.
We pass to Fourier variables
\begin{equation}\label{eq: fourier equation}
    \partial_t\phi_k(t)= -Dk^2\phi_k(t)+ i ak\zeta_k(t),
\end{equation}
where $k \in [-\pi,\pi]$, and $\overline{\zeta_k(s)}=\zeta_{-k}(s)$. The distribution of the noise is Guassian and
$$ \langle\zeta_k(s)\rangle=0 \qquad\langle \zeta_k(s) \zeta_{k'}(s')\rangle=\delta(k+k')\delta(s-s'). $$
The solution of \eqref{eq: fourier equation} is
$$
\phi_k(t)= e^{-t k^2 D}\phi_k(0)+iak\int_{0}^tds e^{-(t-s) k^2 D} \zeta_k(s).
$$
The stationary state of $\phi_k$ is Gaussian with mean zero and covariance
$$\langle \phi_k \phi_{k'} \rangle =-  a^2 k^2 \int_0^t ds e^{-2k^2 D(t-s)}  \delta(k+k') 
= \frac{a^2}{2D}  \delta(k+k') ,$$
which allows to determine $a$, since we know the stationary state in real space, namely
$$
\langle \phi_x \rangle=0, \qquad \langle \phi_x\phi_{x'} \rangle=C\delta({x-x')},
$$
with
$$
C=\sum_i \left( \langle E_0,E_i\rangle- \langle E_0\rangle \langle E_i\rangle \right),
$$
with $E_i$ the microscopic energy densities on the lattice, and  $\langle \cdot ;\cdot\rangle$ the expectation with respect to the microscopic Gibbs state.
Hence, we find the form of the fluctuation-dissipation relation
$$
a^2 =   2DC.
$$
\subsection{Computation of the decorrelation time}
The decorrelation measure $\eta$ for the local fields $\phi_x$ is given by
$$
\eta_x(t)=\frac{\langle (\phi_x(t)-\phi_x(0))^2 \rangle}{2\langle \phi_x^2 \rangle},
$$

\begin{align*}
 &  \langle (\phi_x(t)-\phi_x(0))^2 \rangle = \\
 &\int dk dk'  e^{ix(k+k')} \left(  (e^{tk^2D}-1) (e^{-t{k'}^2D}-1) \langle\phi_k(0) \phi_{k'}(0) \rangle \right.\\  
  & \left. \qquad +
  (ak)^2 \delta(k+k')\int_{0}^t ds  \, e^{-2(t-s) k^2 D}  \right) = \\
 &\int dk dk'  e^{ix(k+k')} \left(  (e^{-tk^2D}-1) (e^{-t{k'}^2D}-1) \frac{a^2}{2D} \delta(k+k') \right. \\  
  & \left. \qquad +
  (ak)^2 \delta(k+k')\int_{0}^t ds  \, e^{-2(t-s) k^2 D}  \right) = \\
   &\int dk   (e^{-tk^2D}-1)^2 \frac{a^2}{2D}   +\int dk
  (ak)^2 \int_{0}^t ds  \, e^{-2(t-s) k^2 D}  .
\end{align*}
The $t\to\infty$ limit of this expression is $2\pi \times 2\times \frac{a^2}{2D}$, which equals $2\langle \phi_x^2 \rangle$.
We therefore get, for large $t$
$$
\langle (\phi_x(t)-\phi_x(0))^2 \rangle-2\langle \phi_x^2 \rangle \approx \frac{a^2}{2D} \left(\frac{1}{\sqrt{2tD}}- \frac{2}{\sqrt{tD}} \right),
$$
and hence
$$
\frac{\langle (\phi_x(t)-\phi_x(0))^2 \rangle}{2\langle \phi_x^2 \rangle} =  1-\frac{1-(2\sqrt{2})^{-1}}{\sqrt{tD}}.
$$

\section{Details for Figure \ref{fig: decorrelation profiles} and \ref{fig: decorrelation time}}\label{Appendix: Figure decorrelation}
The theoretical value $\eta(t)$ defined in \eqref{eq: def eta(t)} involves averages over the thermal ensemble and over disorder. We explain here how we compute this in practice. We approximate
\begin{equation*}
    \eta(t) \approx \frac{1}{D} \sum_{d=1}^D \eta^{(d)}(t)
\end{equation*}
with 
\begin{multline*}
    \eta^{(d)}(t) =\\
    \frac{1}{L} \sum_{k=1}^L \frac{\frac{1}{2}\left( X_k^{(d,d)}(0) - X_k^{(d,d)}(t) \right)^2}{ \frac{1}{N} \sum_{p=1}^N \left( X_k^{(d,p)}(0) \right)^2 - \left( \frac{1}{N} \sum_{p=1}^N X_k^{(d,p)}(0) \right)^2 } .
\end{multline*}
Here $d$ runs over disorder realizations, $k$ runs over all modes and $p$ over initial realizations. 
We use system of size $L=80$, $D=200$ realizations of disorder, and $N=1000$ thermal initial conditions for the denominator.
Notice that in the numerator indices for the realization of disorder and initial conditions coincide, i.e.\@ there is a single initial condition for each disorder realization.
This simply results form the fact that averaging over two sources of randomness, namely the disorder and initial conditions, is naturally done by sampling their joint distribution.
This is not the case in the denominator, because there we are taking the inverse of the thermal average, and it was necessary to vary the initial conditions at a given value of the disorder. 
Luckily, estimating the denominator is numerically cheap since it is a static quantity. 

The simulations were done using the following numerical schemes. 
First, to properly initialize the system, we evolved it form random initial conditions according to Langevin dynamics at inverse temperature $\beta=50$ from time $t=-200$ to time $t=0$. We integrated the Langevin equations using a simple leapfrog method, i.e.\@
\begin{equation}
\left\{\!
\begin{aligned}
    q(t+\tfrac{dt}{2}) &= q(t) + \nabla_p H(q(t),p(t)) \tfrac{dt}{2}\\
    p(t+dt) &= p(t) - \nabla_q H(q(t+\tfrac{dt}{2}),p(t))dt\\
    &+ \left[\xi \nabla_p H(q(t+\tfrac{dt}{2}),p(t)) - r(t)\right]dt,\\
    q(t+dt) &= q(t+\tfrac{dt}{2}) + \nabla_p H(q(t+\tfrac{dt}{2}),p(t+dt)) \tfrac{dt}{2}
\end{aligned}
\right.
\end{equation}
where $r(t) \sim \mathcal{N}\left(0,\sqrt{\tfrac{2\xi}{\beta dt}} \right)$, with coupling strength $\xi = 0.1$ and timestep $dt=0.05$.
We assumed periodic boundary conditions. 
Taking the output of this simulation as thermal initial conditions we continued with the integration of Hamilton equations of motion in the similar manner, i.e.\@ we used the above algorithm with the terms in the third line omitted
\begin{equation}\label{eq: integrator}
\left\{\!
\begin{aligned}
    q(t+\tfrac{dt}{2}) &= q(t) + \nabla_p H(q(t),p(t)) \tfrac{dt}{2}\\
    p(t+dt) &= p(t) - \nabla_q H(q(t+\tfrac{dt}{2}),p(t))dt\\
    q(t+dt) &= q(t+\tfrac{dt}{2}) + \nabla_p H(q(t+\tfrac{dt}{2}),p(t+dt)) \tfrac{dt}{2}.
\end{aligned}
\right.
\end{equation}
We measured the decorrelation level on a numerical grid $t_j = j\cdot 10^s$, $j=0,1,\dots,10^5$ with $s=5$ for $\lambda\leq 6\times 10^{-4}$ and $s=3$ otherwise (we also use $s=0$ for $\lambda\geq 10^{-2}$ for times up to $t=10^4$).
We follow thus the dynamics up to a time $t=10^{10}$ for the smallest values of $\lambda$. 

Decorrelation times where extracted in the following manner. We divided $200$ disorder realization into sets of $20$ realizations over which we averaged $\eta^{(d)}(t)$ obtaining $10$ estimates for $\eta(t)$. For each them we searched for the smallest time $t_j$ such that $\eta(t_j) \geq \eta_*$, where $\eta_*$ is the desired level of decorrelation. With the use of a linear fit to the points $\eta(t_{j-1})$ and $\eta(t_j)$ we extracted the time $t_*$ at which $\eta_*$ is reached. The plotted decorrelation times are means of these $t_*$ and error bars indicate their standard error.

\section{Details for Figure \ref{fig: correlation}}\label{sec: details for figure pretherm}

The blue curve is a numerical estimate for $\text{corr}(p_i,p_{i+1})$ defined with the use of an ergodic average in place of ensemble average.
This correlation was measured after evolving the system of length $L=10^3$
with $\gamma=1$ and fixed $\omega_{L/2}=1$ from the initial condition $p_i=0$, $q_i=c\delta_{i,L/2}$ for $t=10^8$, with $c$ such that the total energy equals $E=0.4$. 
Here we again used the leapfrog method, i.e.\@ we used the algorithm \eqref{eq: integrator} with $dt=0.4$.
The average was taken over a time window $\Delta t = 10^3$ with smaller integration step $dt=0.1$.

The orange curve shows  $\text{corr}(p_i,p_{i+1})$ in the prethermal ensemble constructed from data in a window of size $10$ around each site. More precisely it is computed in a Gibbs ensemble with two conserved quantities $H_0$ and $I$ such that the averages of these quantities are equal to their instantaneous values at time $t=10^8$ in the simulation.

Both curves have been averaged over $100$ realizations of disorder. The black line is the value of $\text{corr}(p_i,p_{i+1})$ in a equilibrium state. It is zero because the equilibrium state is a product for the momentum variables.

\section{Quality of numerical simulations}\label{sec: quality numerics}

\subsection{Length $L$}
\begin{figure}[ht]
\centering
\includegraphics[scale=0.5]{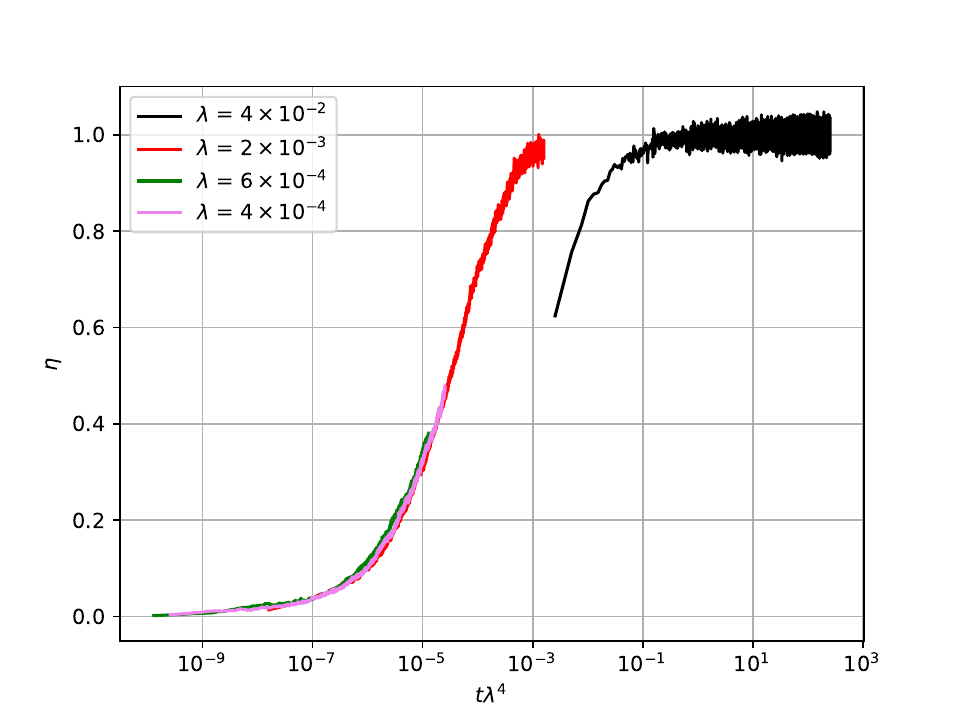}
\caption{\label{fig: eta L=160} The decorrelation $\eta$ as a function of rescaled time $\lambda^4 t$, for a system of length $L=160$. We used here $D=100$ realizations of disorder and measured decorrelation on a numerical grid $t_j = j\cdot 10^s$, $j=0,1,\dots,10^5$ with $s=4$ for $\lambda = 4\times 10^{-4}$ and $s=3$ otherwise. All other parameters were taken the same as for the system of length $L=80$. 
}
\end{figure}

\begin{figure}[ht]
\centering
\includegraphics[scale=0.5]{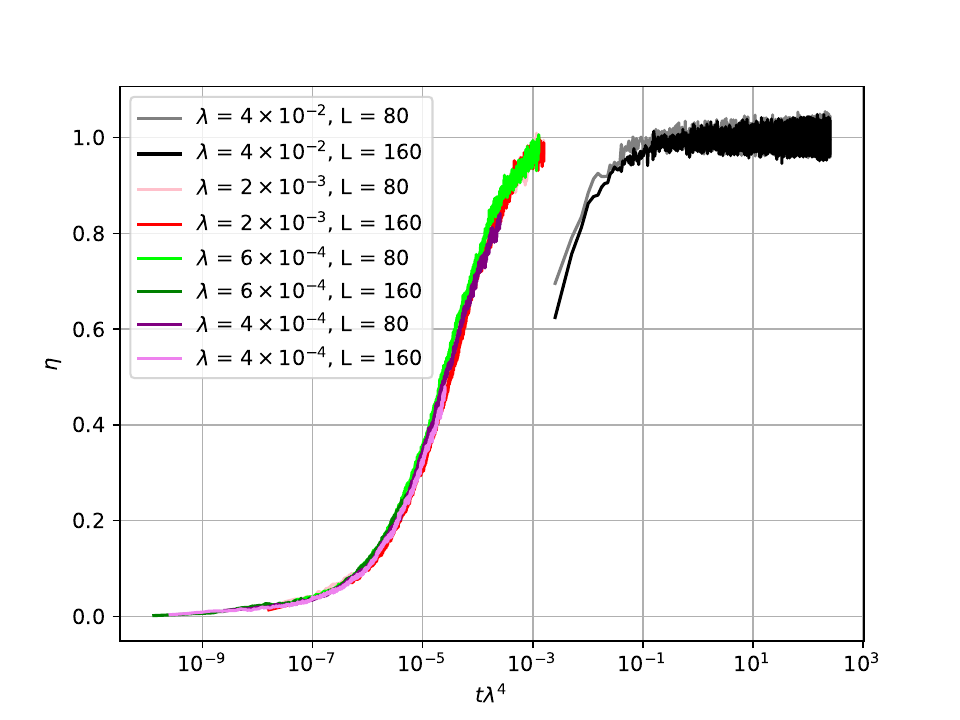}
\caption{\label{fig: eta L=80 and 160} The decorrelation $\eta$ as a function of rescaled time $\lambda^4 t$, for systems of length $L=80$ and $L=160$. We plotted the same curves as in Fig.~\ref{fig: decorrelation profiles} and Fig.~\ref{fig: eta L=160}.
}
\end{figure}

All our equlibrium simulations are performed for a system of size $L=80$.
This may appear rather small as compared to the sizes used in previous works on the spreading of a wave packet, see e.g.~\cite{flach_universal_2009,skokos_delocalization_2009}.
This is however sufficient for a dynamics started from equilibrium with enough disorder averaging. 
To show that our observations are not crucially affected by the system size, we have also probed the behaviour of the system of length $L=160$ for the four largest values of $\lambda$, see Fig.~\ref{fig: eta L=160} and Fig.~\ref{fig: eta L=80 and 160}. 
The data show no significant difference with the data at $L=80$. 

\subsection{Energy conservation}
\begin{figure}[hb]
     \centering
     \begin{subfigure}[b]{0.5\textwidth}
         \centering
         \includegraphics[width=\textwidth]{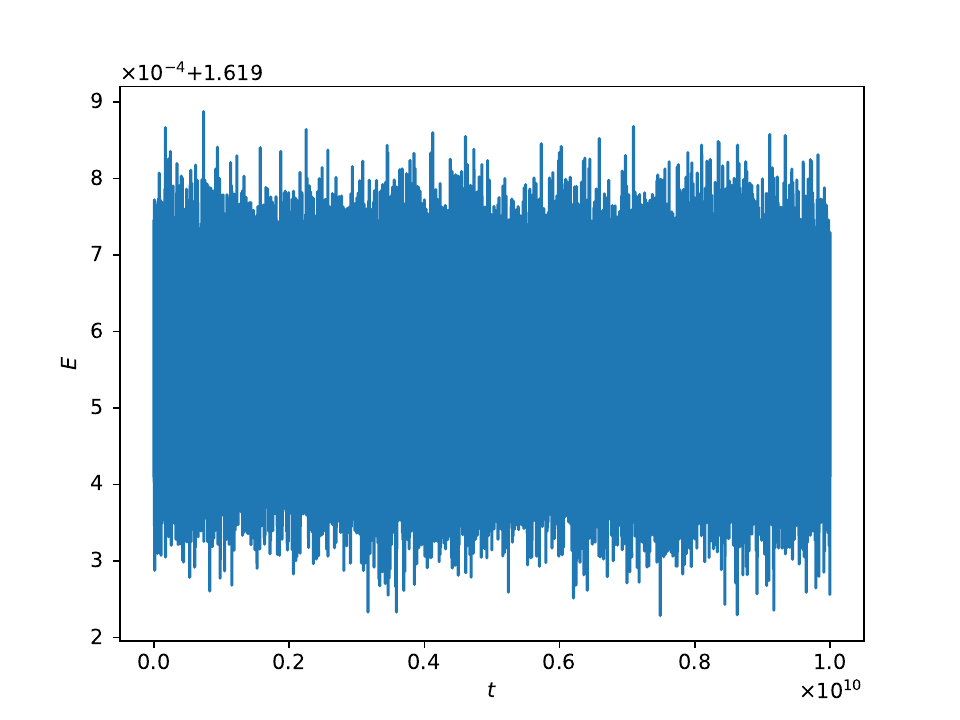}
         \caption{$H(t)$}
     \end{subfigure}
     \begin{subfigure}[b]{0.5\textwidth}
         \centering
         \includegraphics[width=\textwidth]{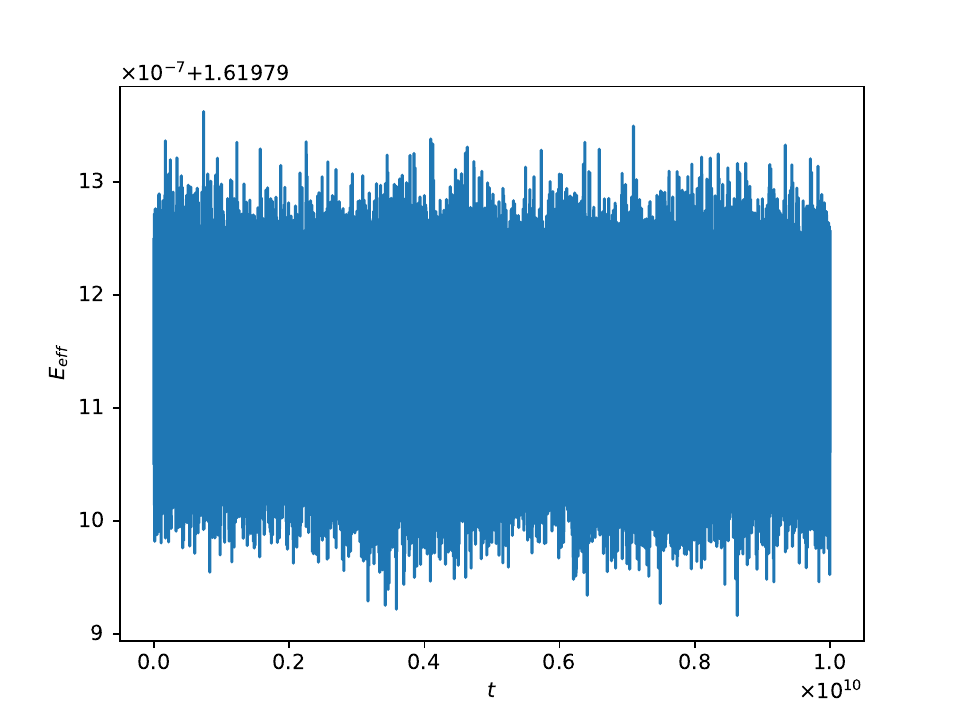}
         \caption{$\Tilde{H}_F(t)$}
     \end{subfigure}
        \caption{Plots of (a) $H(t)$ and (b) $\Tilde{H}_F(t)$ with $\lambda = 2\times 10^{-4}$ evolving from thermal initial conditions as described in Appendix~\ref{Appendix: Figure decorrelation}.}
        \label{fig: H and H_F}
\end{figure}

We comment here on the quality of our numerical integration method. Because of the necessary discretization, the original Hamiltonian $H$ is no longer preserved under dynamics given by \eqref{eq: integrator}. 
However, the scheme \eqref{eq: integrator} is symplectic, 
i.e.\@ it describes the time evolution of a Hamiltonian dynamics, for a time-dependent, periodic, Hamiltonian, and as such can be treated by the Floquet theory.
Henceforth, there exists a time-independent Floquet Hamiltonian $H_F$ which is preserved under the evolution. 
This Floquet Hamiltonian can be expanded in powers of $dt$, and can be shown to take the following form
\begin{equation*}
    H_F = H + dt^2 H_2 + \mathcal{O}(dt^4),
\end{equation*}
where 
\[
H = \sum_{i=1}^L \frac{1}{2} p_i^2 + U(q)
\]
is our original Hamiltonian and where 
\[
H_2 = -\frac{1}{24}\sum_{i,j=1}^L \frac{\partial^2 U}{\partial q_i \partial q_j}p_ip_j + \frac{1}{12}\sum_{i=1}^L \left(\frac{\partial U}{\partial q_i}\right)^2.
\]
In particular, this is consistent with the fact that $H$ remains preserved under the dynamics generated by \eqref{eq: integrator} up to errors of order $\mathcal O(dt^2)$, while the truncated Floquet Hamiltonian $\tilde{H}_F=H + dt^2 H_2$ is preserved up to errors of order $\mathcal{O}(dt^{4})$.
In Fig.~\ref{fig: H and H_F} we provide plots of $H(t)$ and $\Tilde{H}_F(t)$ evolving from thermal initial conditions as described in Appendix~\ref{Appendix: Figure decorrelation}. We observe the expected suppression of the error for $\Tilde{H}_F$, which implies that the round-off errors are well controlled and the main source of error is the discretization in our integration method. 
The relative error is of order $10^{-4}$ for the bare energy and $10^{-6}$ for the truncated Floquet energy, cf.~Fig.~\ref{fig: H and H_F}. More importantly, the error does not noticeably increase with time. 

\section{Extracting the effective anharmonicity from the width $w$}\label{sec: extracting anhar}

We describe here how to assign an effective anharmonicity to a wavepacket of total energy $E$ and width $w$.
If the wavepacket had a box form, with width $\ell$, then the local energy density (which is equal to the temperature at small anharmonicity) is $E/\ell$ and so the effective anharmonicity is obviously
$$
\lambda= \gamma (E/\ell).
$$
Furthermore, a short computation gives the relation between $w$ and $\ell$, namely $\ell= 2\sqrt{3}w$, so that we finally arrive at the estimate
\begin{equation}\label{eq: anharmonicity box}
 \lambda=   \frac{1}{2\sqrt{3}} \gamma    \frac{E}{w}   .
\end{equation}
Actually, assuming a box form is not a bad approximation, as one can see in \cite{tuck_explicit_1976}. It becomes exact for the nonlinear diffusion equation with $D\propto \rho^n$ in the limit of $n\to\infty$. 
The other extreme case is a Gaussian wavepacket, which appears in the opposite limit $n\to 0$. For a Gaussian packet, the energy density is not constant, so we take a weighted average to determine the temperature:
$$
T= \int dx  (1/E)(\rho(x))^2, \qquad \rho(x)= \frac{E}{\sqrt{2\pi}w } e^{-\frac{1}{2}(\frac{x}{w})^2}.
$$
This gives $E/(2\sqrt{\pi}w)$ and so we arrive at
\begin{equation} \label{eq: anharmonicity gauss}
 \lambda=   \frac{1}{2\sqrt{\pi}}\gamma   \frac{E}{w}   .
\end{equation}
Comparing to \eqref{eq: anharmonicity box}, we see that the difference between both estimates is only 3 percent.

\section{Comparison with \cite{vakulchyk_wave_2019}}\label{sec: flach comparison}
In \cite{vakulchyk_wave_2019}, the spreading of a wave packet is investigated numerically for disordered nonlinear discrete-time quantum walks, and the behavior $w\propto t^{1/6}$ is observed for 12 decades in time. 
As eq.~(6) in the SM of \cite{vakulchyk_wave_2019} shows, the equations of motion display a cubic non-linearity at low densities, as in the KG and the DNLS chains. 
All these systems belong thus to the same class, consistently with the obsevations for the spreading of the wave packet. 
Nevertheless, a fair and direct comparison with our numerics is strictly speaking impossible due to the differences between the models. 
Yet, we can estimate the value of the effective non-linearity reached in \cite{vakulchyk_wave_2019} when the packet has maximally spread out.

Reasoning as for the KG chain, we find that the effective non-linearity in the set-up of \cite{vakulchyk_wave_2019} is given by 
$$
    \lambda' = g\rho ,
$$
where $g$ is the bare non-linearity and $\rho$ the density of the packet. 
From the data in \cite{vakulchyk_wave_2019}, we can estimate $\lambda'$ in two ways. 
First, from Fig.~2 in \cite{vakulchyk_wave_2019}, we estimate roughly that $\rho \approx 0.001$ at the final time $t_f = 2 \times 10^{12}$, hence $\lambda' \approx 3\times 10^{-3}$ since $g=3$. 
Second, from Fig.~3 in \cite{vakulchyk_wave_2019}, we also see that at the final time $w^2\approx10^6$. 
Using the procedure described in previous section, we obtain value $\lambda'\approx9\times10^{-4}$.
Either way, despite having run the simulations for much longer times, it is thus not clear to us that the results in \cite{vakulchyk_wave_2019} outperform previous simulations if one considers the effective anharmonicity as the relevant parameter.  

\section{Probing smaller anharmonicities within spreading numerics}\label{sec: going smaller}

From Fig.~{1}. in \cite{flach_universal_2009} we see that $w^2 \approx 900$ is reached at the final time $t_0=10^{10}$ for $E=0.05$. Using the method of Section \ref{sec: extracting anhar}, this corresponds to the value $\lambda_0\approx5\times 10^{-4}$ of the effective anharmonicity. In order to reach the smallest effective anharmonicity considered in our present work, i.e.\ $\lambda_1=2\times 10^{-4}$, under the assumption that the spreading follows the law $w\propto t^{1/6}$,  we estimate that the simulation would need to be continued up to time
\begin{equation}
    t_1 = \left( \frac{\lambda_0}{\lambda_1}\right)^6 t_0.
\end{equation}
Plugging in the numerical values we obtain $t_1\approx 2.5\times 10^{12}$, hence 250 times longer than in \cite{flach_universal_2009}. 
Of course, all estimates of this type should be taken with a grain of salt. Small changes in anharmonicity lead to huge changes in the time.

\end{document}